\documentclass[10pt,pre,onecolumn,aps,showpacs,superscriptaddress,notitlepage,nofootinbib]{revtex4-1}
\usepackage{epsfig,amsmath,amssymb,graphicx,color,calc,epstopdf}
\usepackage{color}
\usepackage{inputenc}

   \let\d=\delta

\def\to{\rightarrow}

\newcommand{\beq}{\begin{equation}}
\newcommand{\eeq}{\end{equation}}
\newcommand{\ba}{\begin{align}}
\newcommand{\ea}{\end{align}}

\makeatletter 
\@addtoreset{equation}{section}
\makeatother  

\usepackage{ifthen,bm}
\newboolean{solutions}
\setboolean{solutions}{false} 
\newcommand{\solution}[1]{\ifthenelse{\boolean{solutions}}{\paragraph{Solution :}#1}{}}
\renewcommand{\vec}{\bm}

\renewcommand{\d}{\mathrm{d}}
\newcommand{\ee}{\mathrm{e}}

\newenvironment{exercisenn}
{\textbf{Exercise:} 
    }
    { 
}
\newenvironment{exercisenb}
{\textbf{Exercise:} 
    }
    { 
}
\newenvironment{exercisebn}
{\textbf{Exercise:} 
    }
    { 
}
\newenvironment{exercise}
{\textbf{Exercise:} 
    }
    { 
}

\usepackage{xr}
\addtocounter{page}{270}

\begin{document}
\title{Lecture Notes on the Statistical Mechanics of Disordered Systems}
\author{Patrick Charbonneau}
\affiliation{Department of Chemistry, Duke University, Durham,
	North Carolina 27708, USA}
\affiliation{Department of Physics, Duke University, Durham,
	North Carolina 27708, USA}

\begin{abstract}
This material complements D.~Chandler's \emph{Introduction to Modern Statistical Mechanics} (Oxford University Press, 1987) in a graduate-level, one-semester course I teach in the Department of Chemistry at Duke University. Students enter this course with some knowledge of statistical thermodynamics and quantum mechanics, usually acquired from undergraduate physical chemistry at the level of D.~A.~McQuarrie \& J.~D.~Simon's \emph{Physical Chemistry: A Molecular Approach} (University Science Books, 1997). These notes, which introduce students to a modern treatment of glassiness and to the replica method, build on the material and problems contained in the eight chapters of Chandler's textbook. 
\end{abstract}


\maketitle
Chapter 5 considered the Ising and related models as a means to explore the consequences of symmetry breaking below the critical temperature, $T_\mathrm{c}$. For $T<T_\mathrm{c}$, the phase space of the Ising model spontaneously breaks in two states of opposite magnetization (or complementary density for a lattice gas, see Fig.~5.4). Because this phenomenology is remarkably common, the approaches used to describe it have been generalized to a broad variety of systems. 

There are, however, a number of models for which symmetry breaking involves more than just a few states. The classical liquids discussed in Chapter 7, for instance, have many disordered local free energy minima upon cooling to $T=0$ (or reaching $p=\infty$ for hard spheres). The ordered (crystal) packing is typically the global minimum, but in systems for which crystal nucleation is slow (compared to the experimental cooling rate) this particular minimum can be avoided, thus resulting the formation of an amorphous solid, i.e., a glass. At low, yet still finite temperature, one can think of the free energy landscape of the system as a collection of free energy basins whose width is related to the number of configurations that share a same energy minimum. The system, however, can eventually jump from one minimum to another, as described in Chapter 8, hence the separation between the basins is not complete.

Various other systems, such as spin glasses, complex networks and even proteins, also contain elements of disorder that seriously hinder their global optimization. This characteristic is in fact observed in disordered systems with a \emph{rugged} free energy landscape, which can have a number of minima that scales exponentially with systems size. The relative number, width and free energy stability of these different states can further change with $T$, even in absence of an external field. These systems thus behave quite differently from the Ising model. But just like simple fluids have a lot in common with the Ising model, a certain universality relates systems with disorder. 

The type of spontaneous symmetry breaking observed in these systems can be described as a loss of \emph{replica symmetry}. The idea is to detect where weakly coupled copies (replicas) of the system start to differ from uncoupled copies. This onset describes a transition between a phase space that is ergodic to one that isn't. We will show below that this approach can also count the number of states and evaluate their properties.

Although the mean-field description of spin glasses we present here is relatively well understood, finite-dimensional corrections to the theory are still somewhat controversial. For instance, while the mean-field description of the Ising model gets remarkably good above dimension $d>4$, where the mean-field values of the critical exponents become exact, similar agreement is only obtained for $d>6$ for the thermodynamic and $d>8$ for the dynamical behavior of spin glasses. And that is only part of the problem. One also needs to account for transitions between various symmetry broken states -- we will briefly get back to this question at the end of the chapter. In spite of these caveats the insights gained from considering the mean-field description of disordered systems amply justifies the effort to study them\footnote{The significance of these advances was recognized by the 1992 Boltzmann and the 1999 Dirac Medals being awarded to Giorgio Parisi for the mean-field solution of spin-glass models. The discovery that spin and structural glasses are formally related at the mean-field level has recently brought renewed attention to this approach.}.

\section{Quenched and Annealed Disorder}

We must first clarify the distinction between annealed and quenched disorder. As a physical example, suppose that we substitute some non-magnetic impurity atoms into a lattice of magnetic ions. For instance, we might mix some fraction of impurities into the melt and crystallize the system by cooling. If this process takes place very slowly, the impurities and the magnetic ions will remain in thermal equilibrium with each other, and the resulting distribution of impurities will have the proper Boltzmann weight for an energy expression that includes the various interactions between the different kinds of atoms. The resulting distribution of impurities is then dubbed \emph{annealed}. If we study systems over the very long time scales necessary to achieve such equilibrium, then we should compute the partition function by not only sampling the possible orientations of the spins of the magnetic ions, but by also sampling over the positions of the impurities. This treatment would be similar to that of Problem 5.27, in which the different realizations of disorder are equilibrated along with the other contributions to the energy. More specifically, given a generalized realization of disorder for a vector of spins $\vec{\sigma}$
\begin{equation}
E_\nu(\mathbf{J},\mathbf{h})=-\vec{\sigma}^{\mathrm{T}}\mathbf{J}\vec{\sigma}-\vec{h}^{\mathrm{T}}\vec{\sigma},
\end{equation}
where $\vec{J}$ and $\vec{h}$ are a matrix and a vector, respectively, with components randomly selected from probability distributions $P_J$ and $P_h$, the partition function of annealed system would be
\begin{eqnarray}
\overline{Q}&=&\int  \d\mathbf{h} \d\mathbf{J} P_h(\mathbf{h})P_J(\mathbf{J}) \sum_{\{\sigma_i=\pm 1\}} \ee^{E_\nu(\vec{J},\vec{h})}\\
&=&\int  \d\mathbf{h} \d\mathbf{J} P_h(\mathbf{h})P_J(\mathbf{J}) Q(\mathbf{J},\mathbf{h}),
\end{eqnarray}
where we denote averaging over all the possible realizations of disorder 
\begin{equation}\label{eq:quenchedavgdef}
\overline{O}=\int  \d\mathbf{h} \d\mathbf{J} P_h(\mathbf{h})P_J(\mathbf{J}) O(\mathbf{J},\mathbf{h}).
\end{equation}

In practice, however, the mobility of impurities in a solid is so small that the timescales involved for them to reach positional thermal equilibrium are beyond human reach. The more common case thus corresponds to regarding the positions of the impurities as fixed, and sampling over only the magnetic degrees of freedom. This is the \emph{quenched} disorder case that we will consider in this chapter. In principle, different realizations of disorder then correspond to distinct systems. Yet for certain quantities, such as the free energy, taking the large system limit is equivalent to averaging over realizations of disorder, and thus corresponds to averaging over the \emph{logarithm} of the partition function,
\begin{equation}
\overline{\ln Q}=\int_{-\infty}^\infty  \d\mathbf{h} \d\mathbf{J} P_h(\mathbf{h})P_J(\mathbf{J}) \ln Q(\mathbf{J},\mathbf{h}).
\end{equation}

\section{Annealed averaging and the replica trick}
Computing the average of the logarithm of a partition function is generally no simple task.
A common way to do so is to rely on the replica trick. More specifically, we first look at $n$ copies of a system with the same realization of disorder $(\mathbf{J},\mathbf{h})$ to obtain
\[
\beta n a(\mathbf{J},\mathbf{h})=-\frac{1}{N}\ln\left[\prod_{\alpha=1}^n \sum_{\{\vec{\sigma}^{(\alpha)}=\pm 1\}} \ee^{-\beta E^{(\alpha)}_\nu(\mathbf{J},\mathbf{h})}\right]=-\frac{1}{N}\ln[Q^n(\mathbf{J},\mathbf{h})].
\]
The replica trick then involves using the following identity
\begin{equation}
\label{eq:reptrick}
\beta \overline{a}=-\frac{1}{N}\overline{\ln Q}=-\frac{1}{N}\lim_{n\rightarrow 0}\frac{\partial}{\partial n}\overline{Q^n}.
\end{equation}
\begin{exercise}
Using the series expansion for an exponential and l'H\^opital's rule, derive the above result.
\solution{
We start from the series expansion for an exponential
\begin{eqnarray}
Q^n&=&\exp(n\ln Q)= (1+n\ln Q+\mathcal{O}(n^2))\\
\lim_{n\rightarrow 0}Q^n&=&\lim_{n\rightarrow 0} (1+n\ln Q)\\
\lim_{n\rightarrow 0}\frac{1}{n}(\overline{Q^n}-1)&=&\overline{\ln Q}\\
\lim_{n\rightarrow 0}\frac{\partial}{\partial n}\overline{Q^n}&=&\overline{\ln Q},
\end{eqnarray}
and use l'H\^opital's rule for the last line.}
\end{exercise}

For a general $n$ this approach is not a simplification of the original problem; however, solving the problem for integer values of $n$ and analytically continuing the result to zero may be. The key difficulty is that this last operation is not mathematically well controlled. Although we use this somewhat uncontrolled trick below, note that rigorous (albeit much more involved) derivations are also available for the systems we consider.

\section{Replica Calculation Example: Fully-connected random-field Ising model (RFIM)}
We first consider a fully-connected system of $N$ ferromagnetically-coupled Ising spins subject to random local fields $h_i$
\begin{equation}
E_\nu(\mathbf{h})=-J\sum_{i,j=1}^N	\sigma_i\sigma_j -\sum_i h_i\sigma_i,
\end{equation}
where the elements of the vector $\mathbf{h}$ are normally distributed, i.e.,
\begin{equation}
P(h_i)\d h_i=\frac{1}{\sqrt{2\pi\sigma_h^2}}\ee^{-h_i^2/(2\sigma_h^2)}\d h_i.
\end{equation}
(Recall that for such a system setting $J=1/N$ guarantees that thermodynamic quantities are extensive, see Appendix.) This mean-field model is known as the random-field Ising model (RFIM), and its average free energy can be obtained by using the replica trick. Following the prescription of the replica trick gives
\begin{eqnarray*}
\overline{Q^n}&=&\int P(\mathbf{h}) Q^n(\mathbf{h}) \d\mathbf{h}\\
&=&\!\!\!\!\sum_{\{\vec{\sigma}^{(\alpha)}=\pm1\}} \!\!\!\!\exp\left[\frac{\beta}{N}\sum_{\alpha=1}^n\sum_{i,j=1}^N \sigma_i^{(\alpha)} \sigma_{j}^{(\alpha)}\right] \int_{-\infty}^{\infty}\frac{\d\mathbf{h}}{(2\pi\sigma_h^2)^{N/2}}\exp\left[-\frac{\mathbf{h}^2}{2\sigma_h^2}+\beta\sum_{\alpha=1}^n \mathbf{h}\cdot \bm{\sigma}^{(\alpha)}\right]\\
&=&\sum_{\{\vec{\sigma}^{(\alpha)}=\pm1\}} \!\!\!\!\exp\left[\frac{\beta}{N}\sum_{\alpha=1}^n\left(\sum_{i=1}^N \sigma_i^{(\alpha)}\right)^2+\frac{\beta^2\sigma_h^2}{2}\sum_{i=1}^N\left(\sum_{\alpha=1}^n \sigma_i^{(\alpha)}\right)^2\right].
\end{eqnarray*}
In order to make progress in simplifying this expression, we use a simple identity for Gaussian integrals known as the Hubbard-Stratonovich transformation
\begin{equation}
\exp \left[ \frac{c y^2}{2} \right]= \frac{1}{(2 \pi c)^{1/2}} \; \int_{-\infty}^{\infty} \exp \left[  \frac{-x^2}{2 c} +  x y \right] \, \d x.
\end{equation}

\begin{exercise}
Verify the validity of the Hubbard-Stratonovich transformation.
\end{exercise}

This transformation allows us to rewrite the first term in the exponential as
\begin{eqnarray*}
& &\prod_{\alpha=1}^{n}\exp\left[\frac{1}{2}\times\frac{1}{2 N\beta}\left(\sum_{i=1}^N 2\beta\sigma_i^{(\alpha)}\right)^2\right]=\\
& &~~~~~~~~~~\left(\frac{N\beta}{\pi}\right)^{n/2}\int \prod_{\alpha=1}^n \d x^{(\alpha)}\exp\left[- N\beta\sum_{\alpha=1}^n (x^{(\alpha)})^2+2\beta\sum_{i=1}^N\sum_{\alpha=1}^n x^{(\alpha)} \sigma_i^{(\alpha)}\right],
\end{eqnarray*}
hence
\begin{eqnarray*}
\overline{Q^n}&=&\left(\frac{N\beta}{\pi}\right)^{n/2}\!\!\!\!\sum_{\{\vec{\sigma}^{(\alpha)}=\pm1\}}\!\!\!\! \int \prod_{\alpha=1}^n \d x^{(\alpha)}\\
& &\times \exp\left[-N\beta\sum_{\alpha=1}^n (x^{(\alpha)})^2+2\beta\sum_{i=1}^N\sum_{\alpha=1}^n x^{(\alpha)} \sigma_i^{(\alpha)}+\frac{\beta^2\sigma_h^2}{2}\sum_{i=1}^N\left(\sum_{\alpha=1}^n \sigma_i^{(\alpha)}\right)^2\right]
\end{eqnarray*}
Note that the exponential is now a simple sum over all $N$ sites, with no interaction between them. The partition sum is therefore the $N$th power of the partition sum of a single site, and we can write
\begin{equation}
\overline{Q^n} = \left(\frac{N\beta}{\pi}\right)^{n/2} \int\prod_{\alpha=1}^n\d x^{(\alpha)}\,\exp\, \left[N\left(-\beta\sum_{\alpha=1}^{n} (x^{(\alpha)})^2 + \ln Q_1(\{x^{(\alpha)}\})\right)\right],
\end{equation}
where 
$Q_1(\{x^{(\alpha)}\})$ is the partition sum for all replicas on a single site
\begin{equation}
Q_1(\{x^{(\alpha)}\}) = \sum_{\{\sigma^{(\alpha)}=\pm 1\}} \exp\left[2\beta\sum_{\alpha=1}^n x^{(\alpha)} \sigma^{(\alpha)}  + \frac{\beta^2\sigma_h^2}{2}\left(\sum_{\alpha=1}^n \sigma^{(\alpha)}\right)^2\right].
\end{equation}
Because the argument of the exponential is proportional to $N$, in the thermodynamic limit, $N\to\infty$, the integral over $x$ is dominated by its maximal term. And because the system is perfectly replica symmetric, that is, all the replicas of the system are exactly equivalent, we also get that $x^{\alpha}$ is the same for all copies at the saddle point, $x^{(\alpha),*}\equiv m$. Taking the derivative of the exponent with respect to $x^{(\alpha)}$ gives an extremum (here, the maximum)
\begin{equation}
-2\beta m+\left.\frac{\partial \ln Q_1(\{x^{(\alpha)}\})}{\partial x^{(\alpha)}}\right|_{x^{(\alpha)}=m}=0,
\end{equation}
and thus
\begin{equation}
m=\frac{1}{Q_1(m)}\sum_{\{\sigma^{(\alpha)}=\pm1\}}\sigma^{(\alpha)} \ee^{\beta a_1[\{\sigma^{(\alpha)}\},m]},
\end{equation}
where 
\begin{eqnarray}
\beta a_1[\{\sigma^{(\alpha)}\},m]&=&2\beta m\sum_{\alpha=1}^n\sigma^{(\alpha)}+ \frac{\beta^2\sigma_h^2}{2}\left(\sum_{\alpha=1}^n \sigma^{(\alpha)}\right)^2\\
\mathrm{and\;\;\;\;}Q_1(m)&=&\sum_{\{\sigma^{(\alpha)}=\pm1\}} \ee^{\beta a_1[\{\sigma^{(\alpha)}\},m]}.
\end{eqnarray}
gives (after dropping the subexponential terms in $n$, which disappear upon taking the limit $n\rightarrow0$)
\begin{equation}
\overline{Q^n}\propto \exp\left[N\left(-n\beta m^2 +\ln Q_1(m)\right)\right].
\end{equation}
Note that the expression for $m$ resembles that for the average magnetization, but for a different Boltzmann weight.

Using the Hubbard-Stratonovich transformation once more, we can write
\begin{equation}
\ee^{\beta a_1[\{\sigma^{(\alpha)}\},m]}=\int \frac{\d x}{\sqrt{2\pi}}\exp\left[-\frac{1}{2}x^2 +(2\beta m+\beta\sigma_hx)\sum_{\alpha=1}^n \sigma^{(\alpha)}\right],
\label{eq:hubstrat2}
\end{equation}
which simplifies the single-site partition function
\begin{eqnarray}
Q_1(m)&=&\int \frac{\d x}{\sqrt{2\pi}}\ee^{-\frac{x^2}{2}} \prod_{\alpha=1}^n\sum_{\sigma^{(\alpha)}=\pm1} \ee^{(2\beta m+\beta\sigma_h x)\sigma^{(\alpha)}}\\
&=&\int \frac{\d x}{\sqrt{2\pi}}\ee^{-\frac{x^2}{2}}[2\cosh(2\beta m+\beta \sigma_h x)]^n\\
&=&\int \frac{\d x}{\sqrt{2\pi}}\exp\left\{-\frac{x^2}{2}+n\ln\left[2\cosh(2\beta m+\beta \sigma_h x)\right]\right\},
\end{eqnarray}
and the self-consistent expression for the magnetization (noting again that the average of the copies is the same as the single-copy average, i.e., it is replica symmetric)
\begin{eqnarray}
m& = &\frac{1}{Q_1(m)}\sum_{\{\sigma^{(\alpha)}= \pm 1\}}\left(\frac{1}{n}\sum_{\alpha=1}^n \sigma^{(\alpha)}\right)\ee^{\beta a_1[\{\sigma^{(\alpha)}\},m]} \\  
&=& \frac{1}{Q_1(m)}\int \frac{\d x}{\sqrt{2\pi}}\, \ee^{-\frac{x^2}{2}} \frac{1}{n}\frac{\partial}{\partial(2\beta m)}\prod_{\alpha=1}^n\sum_{\sigma^{(\alpha)}= \pm 1}\ee^{(2\beta m + \beta \sigma_h x)\sigma^{(\alpha)}}    \\  
&=& \frac{1}{Q_1(m)}\int \frac{\d x}{\sqrt{2\pi}}\, \ee^{-\frac{x^2}{2}} \frac{1}{n}\frac{\partial}{\partial (2\beta m)} \left[2\cosh(2\beta m + \beta \sigma_h x)\right]^n   \\  
&=& \frac{1}{Q_1(m)}\int \frac{\d x}{\sqrt{2\pi}}\, \ee^{-\frac{x^2}{2}+n \ln [2\cosh(2\beta m + \beta \sigma_h x)]} \tanh (2\beta m + \beta \sigma_h x).\label{eq:msce}
\end{eqnarray}
Finally, in order to obtain the disorder-averaged free energy, we use the replica trick as described in Eq.~(\ref{eq:reptrick}),
\begin{eqnarray}
\beta \bar{a}& = &- \frac{1}{N}\left.\frac{\partial \overline{Q^n}}{\partial n}\right|_{n=0} =\beta m^2 -\left.\frac{\partial Q_1(m) }{\partial n}\right|_{n=0} \\ 
& = &\beta m^2 - \int \frac{\d x}{\sqrt{2\pi}}\, \ee^{-\frac{x^2}{2}} \ln \left[2\cosh(2\beta m + \beta \sigma_h x)\right],
\end{eqnarray}
where $m$ is fixed by evaluating the self-consistent equation (\ref{eq:msce}) at $n=0$,
\begin{equation}
m = \int \frac{\d x}{\sqrt{2\pi}}\, \ee^{-\frac{x^2}{2}}\tanh (2\beta m + \beta \sigma_h x).
\end{equation}

\begin{figure}[t]
\begin{center}
\includegraphics[width=0.8\textwidth]{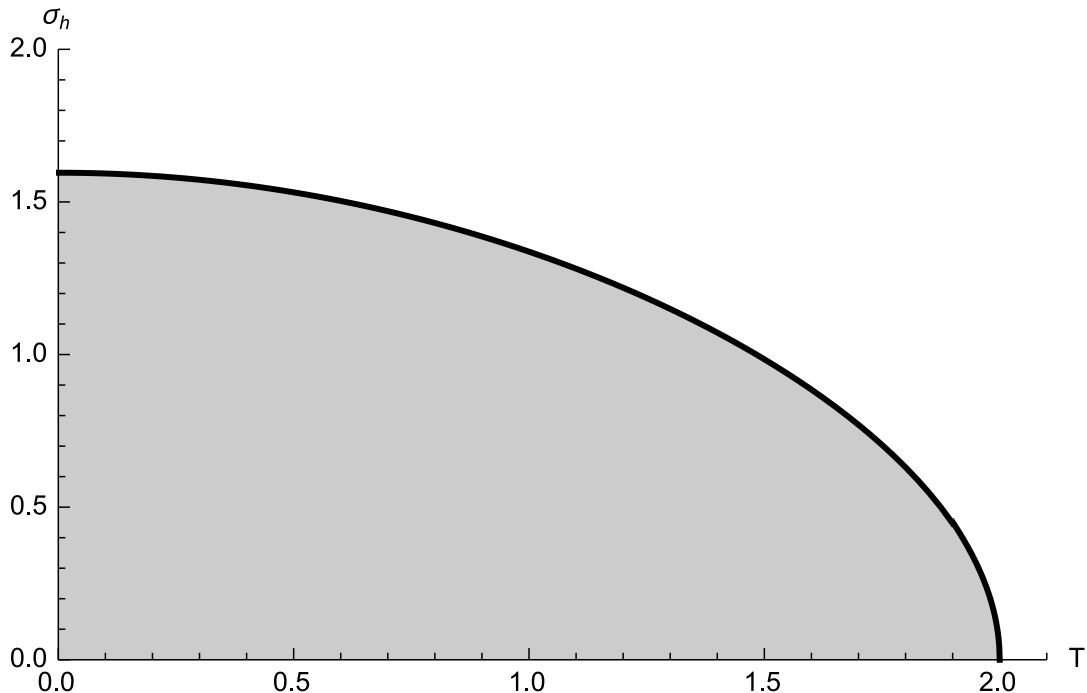}
\end{center}
\caption{Phase diagram of the RFIM model. The ferromagnetic phase (grey zone) is more stable than the paramagnetic phase (white zone) at low temperature and low field $\sigma_h$. A second-order phase transition (black line) separates the two.}
\label{fig:RFIMPD}
\end{figure}

After the change of variables $h\equiv\sigma_h x$, the disorder-averaged free energy becomes
\begin{equation}
\beta \bar{a}= \beta m^2 - \int \frac{\d h}{\sqrt{2\pi\sigma_h^2}}\, \ee^{-\frac{h^2}{2\sigma_h^2}} \ln \left[2 \cosh (2\beta  m + \beta h) \right],
\end{equation}
with
\begin{equation}
m_{\mathrm{SC}}(m)=\int \frac{\d h}{\sqrt{2\pi\sigma_h^2}}\,  \ee^{-\frac{h^2}{2\sigma_h^2}} \tanh \left(2\beta  m + \beta h\right)
\end{equation}
The self-consistent has a trivial solution $m=0$ for all values of $\sigma_h$ and $\beta$. Similarly to what happened in the mean-field treatment of the standard Ising model in Chapter 5, if $\beta$ is sufficiently large, then there exists a non-trivial solution with $m>0$, and thus ferromagnetic order emerges. More specifically, the phase boundary between the paramagnetic and the ferromagnetic phase is found at unit slope
\begin{equation}
2\beta_c \int \frac{\d h}{\sqrt{2\pi\sigma_h^2}}\,  \frac{\ee^{-\frac{h^2}{2\sigma_h^2}}}{\left(\cosh \beta_c h\right)^2} = 1  
\label{eq:RFIMSC}
\end{equation}
The resulting phase diagram is given in Fig.~\ref{fig:RFIMPD}. 
Remarkably, this results indicate the existence of a phase transition driven purely by the fluctuations of the magnetic field, without any thermal noise! The expression for the free energy given above can also be used to deduce more elaborate observables, such as the susceptibility.  For example, near the phase transition the magnetization scales as a power-law,
\begin{equation}
m(T) \sim (T-T_\mathrm{c})^{1/2},
\end{equation}
with the same critical exponent as the mean-field Ising model. This mean-field scaling, however, is here only valid for $d\geq 6$, while the scaling is valid for $d\geq4$ in the standard Ising model. Similarly, the Imry-Ma argument gives that there is no transition in $d=2$ for the RFIM, while the standard Ising model still does. Clearly, magnetic field disorder generally weakens the ferromagnetic ordering transition.

\section{Pure states}
Before considering models with more elaborate types of disorder, we first need to describe \emph{pure states}. (We encountered these states already in Chapter 5, but did not bother to carefully characterize them at the time.) A pure state is a cluster of microstates that become separated from other clusters upon ergodicity breaking in the system. Once pure states emerge, a given copy of the system ends up only accessing a sub-part of phase space through local rearrangements (a properly defined global rearrangement may obviously bring the system from one state to another). In an Ising ferromagnet, for instance, two pure states with $\pm m$ are separated by an infinite barrier in the large system limit in the absence of magnetic field\footnote{In finite dimensions, applying $h\neq0$ allows the nucleation of the lower free energy phase by flipping only a finite number, $\mathcal{O}(N^0)$ of spins, but in infinitely-connected models the barrier is infinite for any $h$.}. Going from one state to the other thus requires flipping $\mathcal{O}(N)$ spins concurrently. 

If the system is decomposed in pure states, its partition function can be expressed as
\begin{equation}
Q=\ee^{-\beta N a(T)}=\sum_{\tilde{\nu}}\ee^{-\beta N a_{\tilde{\nu}}(T)}=\sum_{\tilde{\nu}} Q_{\tilde{\nu}},
\end{equation}
and in order to obtain the average of an observable $O$, we have to calculate
\begin{equation}
\langle O\rangle=\sum_{\tilde{\nu}} w_{\tilde{\nu}} \langle O\rangle_{\tilde{\nu}}\,\,\,\mathrm{with}\,\,\,w_{\tilde{\nu}}=Q_{\tilde{\nu}}/Q
\end{equation}
In the ferromagnetic phase of the Ising model, the average magnetization is indeed
\begin{equation}
m=\frac{1}{2} m_{+}+\frac{1}{2} m_{-}=0,
\end{equation}
because $m_+=-m_-$, by symmetry (see footnote on p.~129). If we have more than a few pure states, however, it is more convenient to rewrite the free energy using a \emph{density of pure states}. Introducing a notation similar to that of the microcanonical ensemble, as described in Section 3.2, we then have that for a free energy distribution, $\tilde{\Omega}(a)$,
\begin{eqnarray}
Q(N,T)&=&\ee^{-\beta N a(T)}=\sum_{\tilde{\nu}}\ee^{-\beta N a_{\tilde{\nu}}(T)}=\int \d a \sum_{\tilde{\nu}} \delta(a-a_{\tilde{\nu}}(T)) \ee^{-\beta N a}\\
&=&\int \d a \tilde{\Omega}(a) \ee^{-\beta N a(T)}=\int_{a_\mathrm{min}}^{a_{\mathrm{max}}} \d a\, \ee^{N[\Sigma(a)-\beta a]}\simeq \ee^{N[\Sigma(a^*)-\beta a^*]},\label{eq:complexity}
\end{eqnarray}
where by analogy to Boltzmann's entropy, we have defined the complexity per spin $\Sigma(a)\equiv\log\tilde{\Omega}(a)/N$. The complexity measures the ``disorder'' of phase space, its configurational entropy so to speak. By construction, $\Sigma\geq0$, and $\Sigma=0$ when the number of pure states grows subexponentially with system size; complexity is then thermodynamically negligible. For the Ising model at $T<T_\mathrm{c}$, for instance, $\Sigma=\ln2/N$, and thus $\Sigma=0$ in the limit $N\rightarrow\infty$. 

The last equality in Eq.~(\ref{eq:complexity}) is obtained from the maximal term method, which we can use in the thermodynamic limit if the complexity is a monotonically increasing function of $a$, i.e., if $\Sigma$ has a single maximum. The partition function can then be approximated by finding the free energy $a^*$ that  maximizes the argument of the exponential, i.e.,
\begin{equation}
\frac{\partial\Sigma}{\partial a}_{a=a^*}=\beta.
\end{equation}

In a glass, unlike in a ferromagnet, states can't be easily separated by symmetry. To identify the transition, we thus use an altogether different strategy. We look for the temperature at which coupled copies of the system become more similar to each other than copies that are not, in the limit of weak coupling. It is similar to selecting a particular magnetization direction in the ferromagnetic Ising model by applying a very weak field in that direction. We will come back to this point later, but the key aspect for now is that we wish to compute the partition function of $m$ copies of the system
\begin{equation}
Q_m=\sum_{\tilde{\nu}} \left(\ee^{-\beta N a_{\tilde{\nu}}}\right)^m=\int_{a_\mathrm{min}}^{a_{\mathrm{max}}} \d a \ee^{N[\Sigma(a)-\beta m a]}\sim \ee^{N[\Sigma(a^*)-\beta m a^*]}, 
\end{equation}
where the saddle point is obtained by setting 
\begin{equation}
\label{eq:mincond}
\left.\frac{\partial\Sigma}{\partial a}\right|_{a=a^*}=\beta m.
\end{equation}
Defining the result of this optimization as 
\begin{equation}\label{eq:phidef}
\Phi(m,T)\equiv m a^*(m,T)-T\Sigma[a^*(m,T)],
\end{equation}
 and assuming that an analytical continuation of $m$ to real values is possible, we obtain
\begin{eqnarray}\label{eq:astar}
a^*(m,T)&=&\frac{\partial \Phi(m,T)}{\partial m}\\
\Sigma[a^*(m,T)]&=&m\beta a^*(m,T)-\beta\Phi(m,T)=m^2\frac{\partial m^{-1}\beta\Phi(m,T)}{\partial m}.
\end{eqnarray}

\begin{exercise}
Check the above identities.
\solution{
\begin{eqnarray}
\frac{\partial \Phi(m,T)}{\partial m}&=&a^* +m\frac{\partial a^*}{\partial m}-T\frac{\partial\Sigma}{\partial a^*}\frac{\partial a^*}{\partial m}\\
&=&a^* +m\frac{\partial a^*}{\partial m}-T\beta m \frac{\partial a^*}{\partial m}\\
&=&a^*
\end{eqnarray}
}
\end{exercise}

From the free energy of $m$ copies of the system in Eq.~(\ref{eq:phidef}), we can thus also obtain the complexity and the optimal free energy of a single copy of the system.

\section{Physical Behavior of the $p$-spin glass model}
As discussed above, adding \emph{quenched disorder} to the coupling constants results in the interaction energy itself being probabilistic\footnote{Note that this interaction energy contrasts with that of glass formation from simple fluids. In this case  disorder arises from the irregular spatial organization of particles, and not from the pair interaction itself. The randomness is then said to be self-generated.}. 
Introducing disorder also weakens phase transitions. It reduces the energy gap at a first-order transition, and can even make a second-order transition disappear altogether, as it does in the RFIM. To increase the probability that a model has a finite-temperature phase transition, we consider the generalized fully-connected $p$-spin models\footnote{Recall that a fully-connected, $p$-spin models undergo a first-order transition for $p\geq3$, while the standard Ising model with $p=2$ has a second-order transition. See Appendix for details.} 
\begin{equation}
E_\nu=-\!\!\!\!\!\!\!\sum_{1\leq i_1<i_2<\ldots<i_p\leq N}\!\!\!\!\!\!\!J_{\mathbf{i}}\sigma_{i_1}\sigma_{i_2}\ldots\sigma_{i_p},
\end{equation}
In order to simplify the analysis, we choose $\sigma_i\in\{-\infty,\infty\}$ to obey the spherical constraint $N=\sum_i\sigma_i^2$, 
and the coupling constants to be normally distributed, i.e.,
\begin{equation}
P(J_{\mathbf{i}})\d J_{\mathbf{i}}=\sqrt{\frac{N^{p-1}}{p!\pi J^2}}\exp\left(-\frac{J_{\mathbf{i}}^2 N^{p-1}}{J^2 p!}\right)\d J_{\mathbf{i}}.
\end{equation}
Note that the distribution properties have been chosen so that the energy per spin remains finite in the  thermodynamic limit $N\rightarrow\infty$.

Before solving this model, let us anticipate the result by first considering the various solution regimes and their microscopic interpretation.
\begin{itemize}
\item For (high) $T>T_\mathrm{d}$, the system is in a disordered paramagnetic phase. The free energy of that phase is smaller than any $a-T\Sigma(a)$ for $a\in[a_{\mathrm{min}},a_\mathrm{max}]$, so the decomposition in pure states is meaningless.
\item For $T_{\mathrm{d}}\geq T\geq T_{\mathrm{K}}$, the paramagnetic state is obtained from the sum of an \emph{exponential number of pure glassy states}, where $a^*\in[a_{\mathrm{min}},a_\mathrm{max}]$. Yet no phase transition occurs at $T_{\mathrm d}$, because the free energy changes smoothly and continuously from that of the paramagnetic phase.  Below $T_{\mathrm d}$, however, the system is dynamically confined to a small region of phase space around where it was initialized. It sits within a pure state. Therefore, $T_{\mathrm d}$ is a dynamical transition, because the paramagnetic state is replaced by a glassy state in which phase space is disconnected, i.e., ergodicity is broken.
\item For (low) $T<T_{\mathrm{K}}$, the system is dominated by the single lowest free energy state $a^*=a_{\mathrm{min}}$ with a complexity $\Sigma(a^*)=0$. At $T_\mathrm{K}$, a phase transition occurs. The free energy and its first derivatives are continuous, but the second derivative jumps, thus giving rise to a second order phase transition, also known as a random first-order phase transition.
\end{itemize}

\section{Replica symmetric calculation of the $p$-spins glass model}
\label{sec:rscacl}
In a replica calculation with replica symmetry breaking, we need $mn$ copies of the system: $m$ copies that will have an infinitely small coupling for calculating the free energy as described above, and $n$ copies for using the replica trick. We thus need to calculate the $n$ replicated free energy of $m$ coupled copies of the system, which have
\begin{eqnarray}
\beta a_m(\mathbf{J}_{\mathbf{i}})&=&-\frac{1}{N}\ln\left[\int' \left(\prod_{\alpha=1}^{m}\d\vec{\sigma}^{(\alpha)}\right) \ee^{-\beta \sum_{\alpha=1}^m E^{(\alpha)}_\nu(\mathbf{J}_{\mathbf{i}})}\ee^{\beta \varepsilon \sum_{\alpha,\alpha'=1}^m\sum_{i=1}^N\sigma_i^{(\alpha)}\sigma_i^{\alpha'}}\right]\nonumber\\
&=&-\frac{1}{N}\ln[Q_m(\mathbf{J}_{\mathbf{i}})]\label{eq:Qm},
\end{eqnarray}
where the prime denotes the spherical constraint on the $\sigma_i$ values. 
Because the coupling terms are infinitely small ($\varepsilon\rightarrow 0$), we drop them from the calculation for now, but they will back later in the analysis.

In the derivation below we drop all non-exponential prefactors because they only provide a trivial offset to the free energy, and thus do not affect the final phase diagram. The partition function is then
\begin{eqnarray}
\overline{Q_m^n}&=&\int' \left(\prod_{\alpha=1}^{mn}\d\vec{\sigma}^{(\alpha)}\right) \!\!\!\!\prod_{1\leq i_1<\ldots<i_p\leq N}\!\!\int \d\mathbf{J}\exp\left[-J_{\mathbf{i}}^2\frac{N^{p-1}}{p!}+\beta J_{\mathbf{i}}\sum_{\alpha=1}^{mn}\sigma_{i_1}^{(\alpha)}\ldots\sigma_{i_p}^{(\alpha)}\right]\nonumber\\
&=&\int' \left(\prod_{\alpha=1}^{mn}\d\vec{\sigma}^{(\alpha)}\right) \!\!\!\!\prod_{1\leq i_1<\ldots<i_p\leq N}\!\!\!\!\exp\left[\frac{\beta^2 p!}{4N^{p-1}}\sum_{\alpha,\alpha'=1}^{mn}\sigma_{i_1}^{(\alpha)}\sigma_{i_1}^{(\alpha')}\ldots\sigma_{i_p}^{(\alpha)}\sigma_{i_p}^{(\alpha')}\right]\nonumber\\
&=&\int' \left(\prod_{\alpha=1}^{mn} \d\vec{\sigma}^{(\alpha)}\right)\exp\left[\frac{\beta^2}{4N^{p-1}}\sum_{\alpha,\alpha'=1}^{mn}\left(\sum_{i=1}^N\sigma_{i}^{(\alpha)}\sigma_{i}^{(\alpha')}\right)^p\right].\label{eq:Qmnbar1}
\end{eqnarray}
Note that the first integral over Gaussian variables simplifies the expression, but couples the replicas, just as it did for the RFIM. Note also that the last line is obtained
by rewriting the product of exponentials into a sum of the exponents.

\subsection{Overlap}
Further simplifying the expression requires getting rid of the constrained integral to rewrite the $p$-tuplets in a more convenient form. In order to do so symmetrically, we introduce an overlap function that measures the similarity between two replicas of the system
\begin{equation}
\chi_{\alpha\alpha'}\equiv\frac{1}{N}\sum_{i}^N\sigma_i^{(\alpha)}\sigma_i^{(\alpha')}.
\end{equation}
For $\alpha=\alpha'$, we obviously get a perfect overlap and recover the spherical constraint, $\chi_{\alpha\alpha'}=1$. For $a\neq b$, the overlap of configurations within a state measures the size of that state in phase space; the larger the overlap, the tighter the cluster, and hence the more similar are the microstates within a pure state. In the low-temperature limit, $T\rightarrow 0$, each pure state concentrates to the lowest energy microstate within the cluster, hence $\chi_{\alpha,\alpha'}\rightarrow1$. As $T$ grows, however, more configurations typically contribute to the pure state, and thus $\chi_{\alpha,\alpha'}<1$. In the (high-temperature) paramagnetic state, the system exhibits no clustering at all, hence all of phase space is accessible and $\chi_{\alpha,\alpha'}=0$.

Using an integral representation for the $\delta$ function, $\delta(x)=\frac{1}{2\pi}\int_{-\infty}^{\infty} \exp{(itx)} \d t$, we can rewrite the overlap as
\begin{eqnarray}
1&=&\int \d\hat{\chi}\prod_{\alpha,\alpha'=1}^{mn}\delta\left(N\chi_{\alpha\alpha'}-\sum_{i}^N\sigma_i^{(\alpha)}\sigma_i^{(\alpha')}\right)\\
&=&\int \d\hat{\chi}\int_{-i\infty}^{i\infty}\d\hat{\lambda}\prod_{\alpha,\alpha'=1}^{mn}\frac{1}{2\pi i}\exp\left[N\lambda_{\alpha\alpha'}\chi_{\alpha\alpha'}-\sum_{i}^N\sigma_i^{(\alpha)}\lambda_{\alpha\alpha'}\sigma_i^{(\alpha')}\right].
\end{eqnarray}

\begin{exercise}
Verify this last result.
\end{exercise}

Plugging this result in Eq.~(\ref{eq:Qmnbar1}) gets rid of the integration constraint, substituting them instead by integrals over $\chi_{\alpha\alpha'}$. We therefore get (again dropping subexponential prefactors)
\begin{eqnarray*}
\overline{Q_m^n}&= &\int \left(\prod_{\alpha=1}^{mn}\d\vec{\sigma}^{(\alpha)}\right) \d\hat{\chi}\d\hat{\lambda}\\
&\times& \exp\left[\frac{\beta^2N}{4}\sum_{\alpha,\alpha'=1}^{mn}\chi_{\alpha\alpha'}^p+N\sum_{\alpha,\alpha'=1}^{mn}\lambda_{\alpha\alpha'}\chi_{\alpha\alpha'}-\sum_{\alpha,\alpha'=1}^{mn}\sum_{i}^N\sigma_i^{(\alpha)}\lambda_{\alpha\alpha'}\sigma_i^{(\alpha')}\right].
\end{eqnarray*}
Using the spectral theorem (as in Section 4.3) decouples the quadratic terms in the exponent. The integrals over the spins then simplify to a multivariate Gaussian integral, hence we can write
\begin{equation}\label{eq:Qmnbar2}
\overline{Q_m^n}=\int \d\hat{\chi}d\hat{\lambda}\exp[N X(\hat{\chi},\hat{\lambda})],
\end{equation}
where we have implicitly defined for matrices $\hat{\chi}$ and $\hat{\lambda}$
\begin{equation}
X(\hat{\chi},\hat{\lambda})\equiv\frac{\beta^2}{4}\sum_{\alpha,\alpha'=1}^{mn}\chi_{\alpha\alpha'}^p+\sum_{\alpha,\alpha'=1}^{mn}\lambda_{\alpha\alpha'}\chi_{\alpha\alpha'}-\frac{1}{2}\ln\det(2\hat{\lambda}).
\end{equation}

\subsection{Saddle-point evaluation}
Equation~(\ref{eq:Qmnbar2}) lends itself to a saddle point approximation in the thermodynamic limit. If the distribution were bimodal (or worse), this option would not be readily available, but the presence of multiple replicas of the system enables us to capture all minima at once. There are a few additional mathematical difficulties. First, the proper limits we need to take are
\begin{equation}
\beta \Phi(m,\beta)=-\lim_{N\rightarrow\infty}\frac{1}{N}\lim_{n\rightarrow 0}\frac{\partial}{\partial n} \int \d\hat{\chi}\d\hat{\lambda}\exp[N X(\hat{\chi},\hat{\lambda})].
\end{equation}
As for the RFIM, in order to make a saddle-point approximation we need to invert the two limits. In general, this operation can be mathematically risky, but it is here (provably) correct.  Second, we must make sure that the optimum of $X$ is not unstable, i.e., that it is not a (high-order) saddle point. All the eigenvalues of the matrix must thus be negative, which in this case can also be shown to be true. Third, in order to optimize the function $X$ with respect to each element of the matrix $\hat{\lambda}$ we need the linear-algebra identity
\begin{equation}
\frac{\partial}{\partial M_{\alpha\alpha'}}\ln\det\hat{M}=(\hat{M}^{-1})_{\alpha'\alpha}.
\end{equation}
\begin{exercise}
Verify the above result.
\end{exercise}

Setting $\left.\partial X/\partial \lambda_{\alpha\alpha'}\right|_{\lambda_{\alpha\alpha'}=\lambda_{\alpha\alpha'}*}=0$ in order to identify the extremum of the exponent then gives
\begin{equation}
\chi_{\alpha\alpha'}-(2\hat{\lambda*}^{-1})_{\alpha'\alpha}=0,
\end{equation}
which upon substitution simplifies the exponent as
\begin{equation}
X(\hat{\chi})=\frac{\beta^2}{4}\sum_{\alpha,\alpha'=1}^{mn}\chi_{\alpha\alpha'}^p+\frac{1}{2}\log(\det\hat{\chi})+\mathcal{O}(nm).
\end{equation}
The leading order correction is linear in $n$, which provides a constant shift to the free energy upon taking the derivative with respect to $n$.  Without loss of generality, we can thus drop it. Higher-order terms in $n$ become exactly zero in the limit $n\rightarrow0$ and can thus also be dropped. The free energy of the replicated system is then
\begin{equation}\label{eq:freeenergy}
\beta \Phi(m,\beta)=-\frac{1}{N}\lim_{n\rightarrow 0}\frac{\partial}{\partial n}\exp \left\{\frac{N}{2}\left[\frac{\beta^2}{2}\sum_{\alpha,\alpha'=1}^{mn}\chi_{\alpha\alpha'}^p+\log(\det \hat{\chi})\right]\right\}.
\end{equation}

\subsection{The Parisi ansatz (enlightened guess)}
We seek a general solution for $\hat{\chi}$, for arbitrary $n$. This seems hard, but cannot be avoided if we are to use the replica trick. As a guess, we postulate that $\hat{\chi}$ takes the form known as the Parisi \emph{ansatz}, which can be motivated as follows. At low temperatures, we expect multiple pure states to be possible, but each of the $m$ weakly coupled replicas, as in Eq.~(\ref{eq:Qm}), should end up in the same state and thus to have an overlap $0<\chi_{\alpha\alpha'}\equiv q\leq1$. By contrast, all the other replicas are overwhelmingly likely to end up in different pure states and thus to have a much lower overlap. (It can be shown that for these replicas $\chi_{\alpha\alpha'}=0$.) If we group replicas by the pure state to which they below, we obtain a matrix of the form
\begin{equation}
\hat{\chi}=\left(
  \begin{array}{cc}
    \left(
       \begin{array}{ccc}
         1 & q & q \\
         q & 1 & q \\
         q & q & 1 \\
       \end{array}
     \right)
     & 0 \\
    0 & \left(
           \begin{array}{ccc}
             1 & q & q \\
             q & 1 & q \\
             q & q & 1 \\
           \end{array}
         \right)
     \\
  \end{array}
\right)
\end{equation}
for $n=2$ sets of $m=3$ copies. We can then use the relation
\begin{equation}
\det   \left(
       \begin{array}{ccc}
         1 & q & q \\
         q & 1 & q \\
         q & q & 1 \\
       \end{array}
     \right)=(1-q)^{m-1}[1+(m-1)q]
\end{equation}
to obtain
\begin{equation}
\det \hat{\chi}=\{(1-q)^{m-1}[1+(m-1)q]\}^n,
\end{equation}
and directly compute
\begin{equation}
\sum_{\alpha\alpha'}^{mn}\chi_{\alpha\alpha'}^p=n[m+m(m-1)q^p].
\end{equation}
Implicitly defining
\begin{eqnarray*}
X(\hat{\chi})&\equiv& -\beta mn\phi_{\mathrm{1RSB}}(m,q,T)\\
&=&\frac{mn}{2}\left\{\frac{\beta^2}{2}[1+(m-1)q^p]+\frac{m-1}{m}\log(1-q)+\frac{1}{m}\log[1+(m-1)q]\right\}.
\end{eqnarray*}
gives an expression that is well defined for any real $n$. We can thus analytically continue it and use the replica trick, which gives the one-step replica symmetry breaking (1RSB) solution\footnote{This result is known as the 1RSB solution, because additional subdivision of phase space are possible, which would correspond to a higher order of replica symmetry breaking. The form of the overall matrix is then adjusted so as to capture this richer state structure.}
\begin{equation}
\beta \Phi(m,\beta)=\frac{1}{N}\lim_{n\rightarrow 0}\frac{\partial}{\partial n}\exp[N\beta mn \phi_{\mathrm{1RSB}}]=m\beta\phi_{\mathrm{1RSB}}(m,q^*,T),
\end{equation}
where $q*$ minimizes the free energy, i.e., its solutions must obey 
\begin{equation}
\left.\frac{\partial\phi(m,q,T)}{\partial q}\right|_{q=q^*}=0 \Rightarrow (m-1)\left(\frac{\beta^2}{2}p q^{p-1}-\frac{q}{(1-q)[(m-1)q+1]}\right)=0.
\end{equation}
Recall that $m$ must be set to unity in the end, because we ultimately care about the free energy of a single replica of the system. These operations must, however, be done with care, because we have analytically continued the free energy expression to unphysical regimes.

\subsection{Phase diagram}
\begin{figure}[t]
\begin{center}
\includegraphics[width=0.8\textwidth]{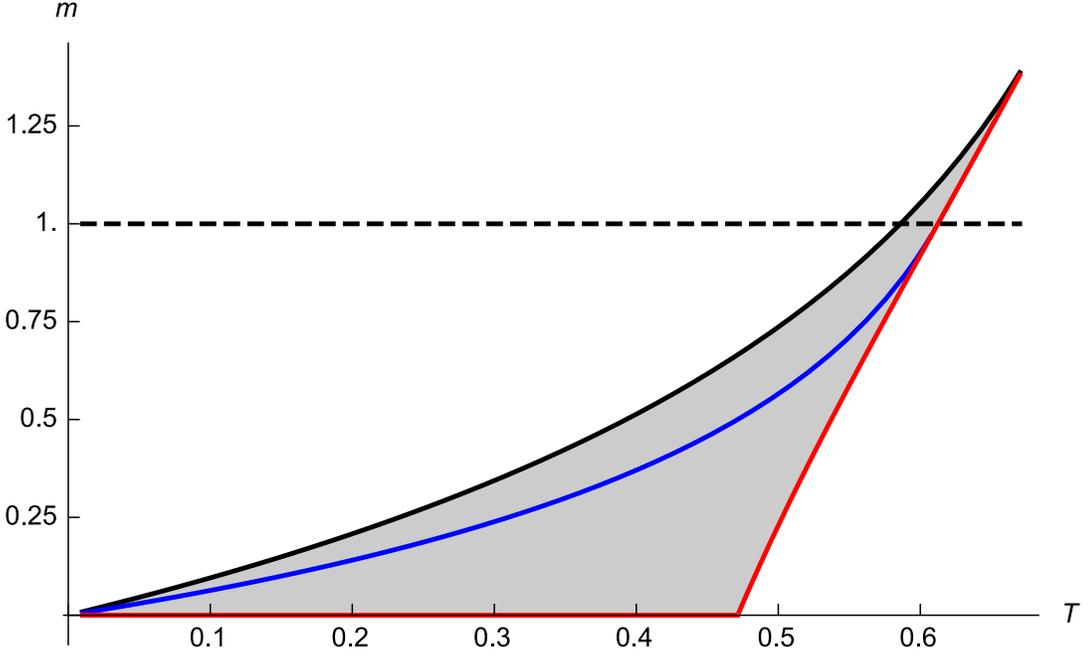}
\end{center}
\caption{Phase diagram of the $3$-spin spherical model in the $m$-$T$ plane. The equilibrium solution is found on the $m=1$ line (dashed). The thermodynamic transition takes place at $T_\mathrm{K}=0.58605$, where $m_\mathrm{s}(T)$ (black line) crosses  the $m=1$ line and $q_\mathrm{K}=0.645$. The overlap is therefore quite significant in the ground state. The onset of breaking up of phase space defines the dynamical transition at $T_\mathrm{d}=0.61237$, where $m_\mathrm{d}(T)$ (blue line) crosses the $m=1$ line and $q_\mathrm{d}=0.5$. The states are then already fairly well-formed. States with $q\neq 0$ are found in the grey zone, above $m^*(T)$ (red line). Adapted from F.~Zamponi, arXiv:1008.4884.}
\label{fig3:dia}
\end{figure}
We have assumed that replicas in different blocks are uncorrelated and have zero overlap, while replicas in
the same block are correlated and thus have an overlap $q$. This correlation is due to the
artificial coupling between the replicas, whose role is only to select the solution for which $q$ has the highest value at any given $T$. As mentioned above, this approach is similar to using an infinitesimal external field to select for one of the two ferromagnetic states of the Ising model. Without this coupling, the free energy from Eq.~(\ref{eq:freeenergy}), which always has a minimum at $q=0$, would not detect the $q^* \neq 0$ stationary points that appear at low $T$.  If we are interested in the partition function of $m$ replicas {\it in the same state},
as we did above to compute the complexity, we should therefore {\it always}
take the solution with $q \neq 0$, if it exists. From the final result
\begin{equation}
\Phi(m,T) = m \phi_\mathrm{1RSB}(m, q^*(m,T),T),
\end{equation}
we can draw a phase diagram in the $m$-$T$ plane. The phase diagram in
Figure 3~\ref{fig3:dia} for $p=3$ results from the following considerations.
\begin{itemize}
\item
We must identify the region where a solution with $q \neq 0$
is found, i.e., where pure states exist. This delimits the line $m^*(T)$. For $m<m^*(T)$, we have that $q^*(m,T)=0$, $\Phi(m,T)$ is trivial, and the complexity is zero; while for $m>m^*(T)$,
a non-trivial solution $q^*(m,T)$ is found. In this region
we can compute the complexity and the free energy 
as a function of $m$ using Eqs.~\ref{eq:astar} from our discussion of pure states
\begin{eqnarray}
\label{eq:SimT}
\Sigma(m,T) &=& m^2 \frac{\partial \left[ m^{-1}\beta \Phi(m,T) \right]}{\partial m} = 
m^2 \frac{\partial [ \beta\phi_\mathrm{1RSB}(m,q^*,T) ]}{\partial m} \ , \\
a^*(m,T) &=& \frac{\partial  \Phi(m,T)}{\partial m} = \frac{\partial m \phi_\mathrm{1RSB}(m,q^*,T)}{\partial m} \ .
\end{eqnarray}
Note that these quantities are {\it only} defined above the $m^*(T)$ line.

\item
Analyzing the behavior of $a^*(m,T)$ and $\Sigma(m,T)$ indicates that starting from $m=\max\{0,m^*(T)\}$ and increasing $m$, both $a^*(m,T)$ and $\Sigma(m,T)$ 
first increase up to a maximum and then decrease upon further increasing $m$. This non-monotonic behavior results in two branches of the
parametric curve $\Sigma(a)$, of which only one is physical: the one that corresponds to both
$a$ and $\Sigma$ {\it decreasing} with $m$, as can be seen from Eq.~(\ref{eq:mincond}).
We can thus define a line $m=m_\mathrm{d}(T)$ where $a^*(m,T)$ is maximum, which corresponds to where the upper threshold values
$a_\mathrm{max}(T)$, i.e., no states of higher free energy exists at that $T$. The relevant region is thus the one for which $m\geq m_\mathrm{d}(T)$.

\item
We observe that
at a fixed $T$, the complexity is finite for $m=m_\mathrm{d}(T)$
and decreases upon increasing $m$ above $m_\mathrm{d}(T)$, until it vanishes on a second line $m_\mathrm{s}(T)$. Above this line, the complexity becomes
negative, indicating that pure states do not exist anymore. States are therefore found for $m_\mathrm{s}(T)>m>m_\mathrm{d}(T)$, with $m_\mathrm{d}(T)$ corresponding to $a_\mathrm{max}(T)$ and
$m_\mathrm{s}(T)$ corresponding to $a_\mathrm{min}(T)$.
\end{itemize}

The line $m=1$ which corresponds to the equilibrium partition function of a single copy of the system. The lines $m_{\mathrm{d}}(T)$ and $m^*(T)$ merge at $m=1$.
When the line $m_\mathrm{d}(T)$ crosses $m=1$, the optimum free energy is $a_\mathrm{max}$. At this point
the paramagnet breaks into many states, and this crossing defines the temperature $T_{\mathrm{d}}$. Similarly, the point where the
line $m_\mathrm{s}(T)$ crosses $m=1$ corresponds to the point where the saddle point is equal to $a_\mathrm{min}$, 
the complexity vanishes and a single state dominates, which corresponds to the 
phase transition at $T_\mathrm{K}$.

For $T<T_\mathrm{K}$, the optimal value of the free energy is always $a_\mathrm{min}$, as obtained along the
$m_\mathrm{s}(T)$. Continuing the free energy above $m_\mathrm{s}(T)$ is incorrect, because the complexity then becomes negative, which is clearly unphysical.
In other words, the system of $m$ replicas undergoes a thermodynamic transition on the line $m_\mathrm{s}(T)$, and the value of the free energy $\Phi(m,T)$ on the line $m_\mathrm{s}(T)$, which is $\Phi(m,T) = m a_{\mathrm{min}}(T)$,
persists up to the $m=1$ line. Noting that $m<1$ and recalling that $\Sigma=0$ along $m_\mathrm{s}$, one obtains from Eq.~(\ref{eq:phidef}) that 
\begin{equation}\label{fglass}
a_\mathrm{min}(T) = \left. \frac{\Phi(m,T)}{ m} \right|_{m=m_\mathrm{s}(T)} = \phi_\mathrm{1RSB}(m_\mathrm{s}(T),q^*(T),T) \ .
\end{equation}
This last result is very important. Indeed, for $T>T_{\mathrm{K}}$ the free energy of the system is equal to the free
energy of the paramagnet, corresponding simply to $\phi_\mathrm{1RSB}(m=1) = -\beta/4$. Below $T_\mathrm{K}$, instead, the free
energy of the system is given by extremizing $\phi_\mathrm{1RSB}$
with respect to {\it both} $m$ and $q$\footnote{Actually, the extremum is here a {\it maximum}. This inversion is characteristic of free energies obtained using the replica trick, and can be shown not to be problematic.}

The above discussion shows that $q$ is the {\it order parameter} for the transition. It is zero in the paramagnetic phase, and it jumps to a nonzero value in the glass phase signaling the spontaneous breaking of replica symmetry.  This assumption is represented in the structure of the overlap matrix. The nature of the low-temperature phase is to have its symmetry spontaneously broken below a certain temperature, with $q$ going from $0$ in the paramagnetic phase, to a finite value at $T_\mathrm{d}$. This behavior is similar to the ferromagnetic $p$-spin mean-field model from the first exercise at the end of this chapter, where the magnetization goes from 0 to a finite value, i.e., the state is well formed at the transition for $p\geq 3$. There is no thermodynamic transition at $T_{\mathrm{d}}$ because the sum over all these pure states is equal to the paramagnetic free energy. Yet the system is dynamically trapped within a pure state, which can be shown explicitly by solving the dynamics of this model (see additional references). By contrast, the second-order transition at $T_\mathrm{K}$ \emph{is} thermodynamic, because the free energy then behaves differently from that of the paramagnetic phase. 

It can be shown that in general the dynamical transition takes place at
$T_\mathrm{d}=\sqrt{\frac{p(p-2)^{p-2}}{2(p-1)^{p-1}}}$ with
$q_\mathrm{d}=\frac{p-2}{p-1}$,
while 
$T_{\mathrm{K}}=\sqrt{\frac{py^*}{2}}(1-y^*)^{1-2/p}$ with 
$q_{\mathrm{K}}=1-y^*$, where $y^*$ is the solution to
$\frac{2}{p}=-2y\frac{1-y+\ln y}{(1-y)^2}$.
Note that both transitions weaken as $p$ disappear and vanish for $p=2$. This effect is reminiscent of what happens in systems without disorder.

In summary, the replica method allow us to fully characterize the thermodynamics of the spherical $p$-spin
model, by computing the free energy of the paramagnetic phase, the free energy of the glass phase,
and the distribution $\Sigma(a)$ of all the metastable states as a function of $a$ and $T$.
The main assumption we made in the derivation is that {\it all the states are equivalent}, with a
self-overlap $q$ and zero mutual overlap, which suggests that they are randomly distributed in phase space. 
These properties are expressed
by the 1RSB structure of the overlap matrix $\chi_{\alpha\alpha'}$, in which all nonzero entries, correspond to replicas evolving in the same pure state, have an equivalent overlap $q$.
This behavior
is exact for the spherical $p$-spin glass model, but this is somewhat exceptional. The general description of the transitions is nonetheless quite general. 


\section{Finite $d$ corrections}
A description of some of the finite-dimensional corrections to the mean-field picture due is contained within the random first-order transition (RFOT) theory of the (spin-)glasses initially formulated by T.~Kirkpatrick, D.~Thirumalai, and P.~Wolynes in the 1980s. RFOT notably suggests that  barriers between states remain finite if the model is considered on a finite-dimensional lattice. Going from one state to another is then possible via the nucleation of a lower free energy state. The barriers then result from the competition between the free energy difference and the interfacial free energy between the various states. Because the scaling of these two quantities converges in high dimension, we recover that the barrier heights diverge in the infinitely-connected limit, so pure states are then well defined. The role of fluctuations in low-dimensions is also non negligible, and adding them to the mean-field description is still an active area of research. In structural glasses, excitations due to local structural defects also affect the system dynamics. This last effect was extensively considered by David Chandler and his collaborators in the 2000s.
\begin{acknowledgments}
I am grateful to the many former students and collaborators who have helped me improve these notes. I am especially indebted to Francesco Zamponi, who has patiently introduced me to these ideas. This work was supported in part by a grant from the Simons 
Foundation (\#454937) and by National Science Foundation grant no. NSF DMR-1055586.
\end{acknowledgments}
\appendix
\section{Generalized, fully-connected spin models}
In this Appendix, we consider generalized versions of the fully-connected Ising model that are useful for the mean-field study of spin glasses.
Consider a system of $N$ spins that are all nearest-neighbors to each other. This arrangement is the fully-connected (or infinite-dimensional) version of the lattice depicted in Fig.~5.1. Note that for such a system the mean-field solution is exact because nearest-neighbor fluctuations have but a negligible impact on the effective field felt by a particle. 

We also generalize the Ising model by including interactions between multiple sites. The $p$-spin version of the Ising model has spin variables $\sigma=\pm1$ coupled in (integer) $p$-plets (pairs for $p=2$, triplets for $p=3$, etc.), 
\begin{equation}
E_\nu=-J\!\!\!\!\!\!\!\!\!\!\!\sum_{1\leq i_1< i_2<\ldots <i_p\leq N}\!\!\!\!\!\!\!\!\!\!\! \sigma_{i_1}\sigma_{i_2}\ldots\sigma_{i_p}
\end{equation}
where the choice of coupling constant $J\equiv p!/N^{p-1}$ guarantees that the energy per spin remains finite in the thermodynamic limit $N\rightarrow\infty$. 

\begin{exercisebn}
\label{first exercise}
Use mean-field arguments to show that the free energy of the Ising $p$-spin model in the limit $N\rightarrow\infty$ is given by (setting $J=1$)
\begin{equation}
\beta a\equiv\lim_{N\rightarrow\infty}\frac{\beta A}{N}=-\beta m^p+\left[\left(\frac{1+m}{2}\right)\log\left(\frac{1+m}{2}\right)+\left(\frac{1-m}{2}\right)\log\left(\frac{1-m}{2}\right)\right],
\end{equation}
where $m=\langle M\rangle/N=\frac{1}{N}\sum_{i=1}^N \sigma_i$ is the average magnetization per spin. Hint: Write the expression starting from $A=E-TS$.

\solution{For the entropy contribution we can use a microcanonical-like treatment to obtain that the total number of ways to realize a magnetization $m$. We need to count how many ways there are to place $N(1+m)/2$ spins up and $N(1-m)/2$ spins down.
\begin{equation}
\Omega(m,N)=\left(\begin{array}{c}
N\\
N(1+m)/2
\end{array}\right)=\frac{N!}{(N(1+m)/2)!(N(1-m)/2)!}
\end{equation}
We therefore get that
\begin{eqnarray}
S_N&=&\ln\Omega_N(m)\\
&=&N\ln N-N-N\left(\frac{1+m}{2}\right)\left[\ln\left(\frac{1+m}{2}\right)+\ln N-1\right]-N\left(\frac{1-m}{2}\right)\left[\ln\left(\frac{1-m}{2}\right)+\ln N-1\right]\\
&=&-N\left[\left(\frac{1+m}{2}\right)\ln\left(\frac{1+m}{2}\right)+\left(\frac{1-m}{2}\right)\ln\left(\frac{1-m}{2}\right)\right]
\end{eqnarray}.

For the energy, we note that all of the terms now have
\begin{equation}
\sigma_{i_1}\sigma_{i_1}\ldots\sigma_{i_p}=m^p.
\end{equation}
In the limit where $N\gg p$, we have that the total number of terms in the sum $N!/(N-p)!p!\approx N^p/p!$ and therefore the energy contribution follows.
}
\end{exercisebn}
\begin{exercisenn}
In order to better grasp the behavior of the above model, find the series expansion of $a$ to order $m^4$ for $p=2$, and plot the original $a$ as a function of $m$ for $\beta=5/6,1,2,3$, and $4$ for both $p=2$ and $p=4$. What are the equilibrium states?  For $p=4$, if the system is prepared in a metastable state, how long will the system remain in that state?

\solution{
By symmetry, only even terms can appear
\begin{equation}
\beta a=-\ln2+(1-\beta J)\frac{m^2}{2}+\frac{m^4}{12}+\mathcal{O}(m^6).
\end{equation}

\begin{figure}
\begin{center}
\includegraphics[width=0.3\textwidth]{PS5_2spin.eps}\includegraphics[width=0.3\textwidth]{PS5_4spin.eps}
\end{center}
\caption{$p=2$ and $p=4$ free energy density for various temperatures, increasing from top to bottom.}
\label{fig:pspin}
\end{figure}
}
\end{exercisenn}

\begin{exercisenn} By taking the derivative of the Ising $p$-spin model free energy obtained above with respect to the magnetization, show that the self-consistent mean-field relation for the $p=2$ Ising model is recovered. Find the numerical value of the transition temperature and plot the equilibrium magnetization as a function of temperature for both $p=2$ and $p=4$. What is different between the two systems?

\solution{By taking the derivative we obtain
\begin{equation}
\frac{\partial a}{\partial m}=-\beta m^{p-1}+\frac{1}{2}\ln\left(\frac{1+m}{1-m}\right)=0,
\end{equation}
which gives
\begin{equation}
m=\tanh(\beta m^{p-1})
\end{equation}
and is very similar to the mean-field Ising criterion when $p=2$.

For the $p=2$ system a second-order transition is obtained at $T=1$, while for $p=4$ a first-order transition is obtained at $T=0.362949$. Because the free energy barriers to go from one phase to the other grow with the system size, metastable states will persist for an infinite time in the thermodynamic limit.

\begin{figure}
\begin{center}
\includegraphics[width=0.3\textwidth]{PS5_2spin_m.eps}\includegraphics[width=0.3\textwidth]{PS5_4spin_m.eps}
\end{center}
\caption{$p=2$ and $p=4$ equilibrium magnetization as function of temperature.}
\label{fig:pspinm}
\end{figure}}
\end{exercisenn}

\begin{exercisenb} By looking at the second derivative of the free energy, show that the $p=2$ behavior is qualitatively different than for $p\geq3$.

\solution{Taking the derivative of the result from the previous part, we get
\begin{equation}
\frac{\partial }{\partial m}\left[-\beta m^{p-1}+\frac{1}{2}\ln\left(\frac{1+m}{1-m}\right)\right]=-\beta(p-1)m^{p-2}+\frac{1}{1-m^2},
\end{equation}
which gives a critical behavior when $p=2$, but not for all other values of $p$. The transition is thus quite different.

}
\end{exercisenb}

For mathematical convenience we also consider the \emph{spherical} $p$-spin model, where the spin variable can take any real value, i.e., $\sigma_i\in (-\infty,\infty)$, under the constraint
\begin{equation}\label{eq:constraint}
\frac{1}{N}\sum_i \sigma_i^2 =1.
\end{equation}
Note that this condition is implicitly included in Ising models.

\begin{exercise}
We will now solve the mean-field problem for the $p$-spin \emph{spherical} model. Note that in this case, the mean-field energy is also
\begin{equation}
\langle E\rangle=-Nm(\sigma)^p,
\end{equation}
but the entropy expression cannot be written down as straightforwardly as before. We write the partition function as (setting $k_B=1$)

\begin{eqnarray*}
Q &=& \int \d\vec{\sigma} \delta\left(\sum_i \sigma_i^2 - N\right) \ee^{\beta N m(\vec{\sigma})^p}\\
&=& \int \d m \d\vec{\sigma} \delta\left(\sum_i \sigma_i^2 - N\right) \delta\left(\sum_i \sigma_i - N m\right) \ee^{\beta N m^p}\\
&=& \int \d m \ee^{N s(m) + \beta N m^p},
\end{eqnarray*}
where
\begin{equation}
\ee^{N s(m)} = \int \d\vec{\sigma} \delta\left(\sum_i \sigma_i^2 - N\right) \delta\left(\sum_i \sigma_i - N m\right).
\end{equation}
Rewrite this expression using the integral representation for the delta function, $\delta(x)=\int_{-i\infty}^{i\infty} \frac{\d y}{2\pi} \ee^{x y}$, before using a saddle point approximation (for $N\rightarrow\infty$) to calculate the resulting integral and obtain
\begin{equation}
\beta a\equiv\lim_{N\rightarrow\infty}\frac{\beta A(m)}{N} = -\frac{1}{2}\left(1+\ln(2)+\ln(1-m^2)\right)- \beta m^p.
\end{equation}

\solution{Neglecting pre-exponential constant factors, we get
\begin{eqnarray}
\ee^{N s(m)}&=\int_{-i\infty}^{i\infty} d\lambda_1 \int_{-i\infty}^{i\infty} \d\lambda_2 \int \d\vec{\sigma} \ee^{ -\lambda_1\sum_i \sigma_i^2 + N\lambda_1-\lambda_2\sum_i\sigma_i+\lambda_2 Nm}\\
&=\int_{-i\infty}^{i\infty} \d\lambda_1 \int_{-i\infty}^{i\infty} \d\lambda_2 \ee^{N\left(\frac{\lambda_2^2}{4\lambda_1}-\frac{1}{2}\ln\lambda_1+\lambda1+m\lambda_2\right)}.
\end{eqnarray}
Taking the saddle point approximation, we find that
\begin{eqnarray}
\lambda_1&=&\frac{1}{2(1-m^2)}\\
\lambda_2&=&\frac{m}{m^2-1}
\end{eqnarray}
and thus
\begin{equation}
s(m)=\frac{N}{2}\left(1+\ln(2)+\ln(1-m^2)\right),
\end{equation}
which upon substitution gives the proper free energy expression.
}
\end{exercise}

\section*{Bibliography}
Various additional pedagogical references about the $p$-spin model are available. Part of these notes borrowed heavily from the first one.
\begin{itemize}
\item F.~Zamponi, ``Mean field theory of spin glasses'', arXiv:1008.4844.
\item T.~Castellani and A.~Cavagna, ``Spin-glass theory for pedestrians'', \emph{J. Stat. Mech.} P05012 (2005).
\item V.~Dotsenko, \emph{Introduction to the Replica Theory of Disordered Statistical Systems}, Cambridge University Press, Cambridge (2000).
\end{itemize}
Solving the dynamical behavior of the spherical $p$-spin glass is beyond the scope of this chapter, but a pedagogical introduction can be found in Castellani and Cavagna's notes.

Applications of spin glass ideas to protein folding are reviewed in the following.
\begin{itemize}
\item J.~N.~Onuchic, Z.~Luthey-Schulten, and P.~G.~Wolynes, ``Theory of protein
folding: the energy landscape perspective'', Annu.~Rev.~Phys.~Chem. \textbf{48}, 545 (1997).
\end{itemize}
The connection of the above ideas with the mean-field description of structural glasses was also recently reviewed.
\begin{itemize}
\item P.~Charbonneau, J.~Kurchan, G.~Parisi, P.~Urbani, F.~Zamponi, ``Glass and Jamming Transitions: From Exact Results to Finite-Dimensional Descriptions'', Annu.~Rev.~Cond.~Matt.~Phys. \textbf{8}, 265 (2017).
\end{itemize}

Applications of spin glass ideas to computer science and information theory are presented in the following textbooks.
\begin{itemize}
\item M.~M\'ezard and A.~Montanari, \emph{Information, Physics, and Computation}, Oxford University Press, Oxford (2009).
\item H.~Nishimoro, \emph{Statistical Physics of Spin Glasses and Information Processing}, Oxford University Press, Oxford (2001).
\end{itemize}

\end{document}